\documentclass[fleqn,usenatbib]{mnras}
\usepackage{newtxtext,newtxmath}
\usepackage[T1]{fontenc}

\DeclareRobustCommand{\VAN}[3]{#2}
\let\VANthebibliography\thebibliography
\def\thebibliography{\DeclareRobustCommand{\VAN}[3]{##3}\VANthebibliography}

\usepackage{graphicx}	% Including figure files
\usepackage{amsmath}	% Advanced maths commands
\usepackage{xspace}
\usepackage{xcolor}
\usepackage{soul}
\newcommand{\Msun}{\ensuremath{\mathrm{M}_\odot}\xspace}
\newcommand{\kms}{\ensuremath{\mathrm{km\,s}^{-1}}\xspace}
\newcommand{\lamr}{\ensuremath{\lambda_{R}}\xspace}
\newcommand{\lamre}{\ensuremath{\lambda_{R_\mathrm{e}}}\xspace}
\newcommand{\vosigma}{\ensuremath{v/\sigma}\xspace}
\newcommand{\vosigmae}{\ensuremath{(v/\sigma)_\mathrm{e}}\xspace}
\newcommand{\epsilone}{\ensuremath{\epsilon_{\mathrm{e}}}\xspace}
\newcommand{\nsample}{\ensuremath{51}\xspace}
\newcommand{\srfracN}{\ensuremath{9}\xspace}
\newcommand{\srfrac}{\ensuremath{18}\xspace}
\newcommand{\pvalue}{\ensuremath{93}\xspace}
\newcommand{\mangasrfrac}{\ensuremath{19}\xspace}
\newcommand{\mangasrfracstd}{\ensuremath{6}\xspace}

\title[MAGPI spin-ellipticity distribution]{The MAGPI Survey: Massive slow rotator population in place by $z \sim 0.3$}

\author[Derkenne et al.]{
Caro Derkenne,$^{1,2}$
Richard M. McDermid,$^{1,2}$\thanks{E-mail:richard.mcdermid@mq.edu.au}
Francesco D'Eugenio,$^{3,4}$
Caroline Foster,$^{2,5}$
\newauthor
Aman Khalid,$^{2,5}$
Katherine E. Harborne,$^{2,6}$
Jesse van de Sande$^{2,7}$,
Scott M. Croom,$^{2,7}$,
\newauthor
Claudia D.P. Lagos,$^{2,6}$ 
Sabine Bellstedt,$^{6}$ 
J. Trevor Mendel,$^{2,8}$
Marcie Mun$^{2,8}$
\newauthor
Emily Wisnioski,$^{2,8}$
Ryan S. Bagge$^{2,5}$
Andrew J. Battisti,$^{2,8}$
Joss Bland-Hawthorn,$^{7}$
\newauthor
Anna Ferr\'{e}-Mateu,$^{9,10,11}$
Yingjie Peng$^{12,13}$
Giulia Santucci,$^{2,6}$
Sarah M. Sweet$^{2,14}$,
\newauthor
Sabine Thater$^{15}$,
Lucas M. Valenzuela$^{16}$,
and Bodo Ziegler$^{15}$
\\
$^{1}$Astrophysics and Space Technologies Research Centre, School of Mathematical and Physical Sciences, Macquarie University,
NSW 2109, Australia\\
$^{2}$ARC Centre of Excellence for All Sky Astrophysics in 3 Dimensions (ASTRO 3D), Australia\\
$^{3}$Kavli Institute for Cosmology, University of Cambridge, Madingley Road, Cambridge, CB3 0HA, United Kingdom\\
$^{4}$Cavendish Laboratory - Astrophysics Group, University of Cambridge, 19 JJ Thomson Avenue, Cambridge, CB3 0HE, United Kingdom\\
$^{5}$School of Physics, University of New South Wales, Sydney, NSW 2052, Australia\\
$^{6}$International Centre for Radio Astronomy Research, The University of Western Australia, 35 Stirling Highway, Crawley, WA 6009, Australia\\
$^{7}$Sydney Institute for Astronomy, School of Physics, University of Sydney, NSW 2006, Australia\\
$^{8}$ Research School of Astronomy and Astrophysics, Australian National University, Canberra, ACT 2611, Australia\\
$^{9}$Instituto de Astrof\'isica de Canarias, V\'ia L\'actea s/n, E-38205 La Laguna, Tenerife, Spain\\
$^{10}$Departamento de Astrofisica, Universidad de La Laguna, E-38200, La Laguna, Tenerife, Spain\\
$^{11}$Center for Astrophysics and Supercomputing, Swinburne University, John Street, Hawthorn, VIC 3122, Australia\\
$^{12}$Department of Astronomy, School of Physics, Peking University, 5 Yiheyuan Road, Beijing 100871, People’s Republic of China\\
$^{13}$Kavli Institute for Astronomy and Astrophysics, Peking University, 5 Yiheyuan Road, Beijing 100871, People’s Republic of China\\
$^{14}$School of Mathematics and Physics, University of Queensland, Brisbane, QLD 4072, Australia\\
$^{15}$Department of Astrophysics, University of Vienna, T\"{u}rkenschanzstrasse 17, 1180 Wien, Austria\\
$^{16}$Universit\"{a}ts-Sternwarte, Fakult\"{a}t f\"{u}r Physik, Ludwig-Maximilians-Universit\"{a}t M\"{u}nchen, Scheinerstr. 1, 81679 M\"{u}nchen, Germany\\
}

\date{Accepted XXX. Received YYY; in original form ZZZ}

\pubyear{2024}

\begin{document}
\label{firstpage}
\pagerange{\pageref{firstpage}--\pageref{lastpage}}
\maketitle

% Abstract of the paper
\begin{abstract}

We use the `Middle Ages Galaxy Properties with Integral field spectroscopy' (MAGPI) survey to investigate whether galaxies have evolved in the distribution of their stellar angular momentum in the past 3-4 Gyr, as probed by the observational proxy for spin, \lamr. We use 2D stellar kinematics to measure \lamr along with detailed photometric models to estimate galaxy ellipticity. The combination of these measurements quantifies the kinematic classes of `fast rotators' and the rarer `slow rotators', which show no regular rotation in their line-of-sight velocity fields. We compare \nsample MAGPI galaxies with $\log_{10} (M_{\star}/\Msun) > 10$ to carefully drawn samples of MaNGA galaxies in the local Universe, selected to represent possible descendants of the MAGPI progenitors. The EAGLE simulations are used to identify possible evolutionary pathways between the two samples, explicitly accounting for progenitor bias in our results and the varied evolutionary pathways a galaxy might take between the two epochs. We find that the occurrence of slow rotating galaxies is unchanged between the MAGPI ($z \sim 0.3$) and MaNGA ($z \sim 0$) samples, suggesting the massive slow rotator population was already in place $\sim 4$ Gyr ago and has not accumulated since. There is a hint of the MAGPI sample having an excess of high \lamr galaxies compared to the MaNGA sample, corresponding to more ordered rotation, but statistically the samples are not significantly different. The large-scale stellar kinematics, as quantified through the \lamr parameter, of galaxies at $z \sim 0.3$ have already evolved into the diversity of structures seen today in the local Universe.

\end{abstract}

% Select between one and six entries from the list of approved keywords.
% Don't make up new ones.

%%%%%%%%%%%%%%%%%%%%%%%%%%%%%%%%%%%%%%%%%%%%%%%%%%

%%%%%%%%%%%%%%%%% BODY OF PAPER %%%%%%%%%%%%%%%%%%

\section{Introduction}
\label{sec:introduction}

Integral field spectroscopy (IFS) allows us to observe a wealth of detail in the kinematic structure of galaxies, unlocking their histories through their stellar motions \citep{cappellar_2016_structure}. Of particular interest is assessing the relative amounts of ordered and random motions within a galaxy, as this kinematic signature can encode the various accretion and merger events a galaxy has experienced \citep{Naab_2014_analysis,lagos_2018_spin}.  
One method of quantifying the relative amount of pressure and rotation support in a galaxy is to measure the ratio between the line-of-sight velocity and the line-of-sight velocity dispersion, \vosigma \citep{binney_1978_rotation}. In the context of IFS stellar kinematics this is given as \citep[][eq. 26]{binney_2005_rotation}: 
\begin{equation}
    v/\sigma = \sqrt{\frac{\sum_{i=1}^{N}F_iv_i^2}{\sum_{i=1}^{N}F_i\sigma_i^2}},
	\label{eq:vsigma}
\end{equation}
where $N$ is the number of spaxels in the chosen aperture, $v_i$ is the line-of-sight velocity of a spaxel measured relative to the average line-of-sight velocity of the galaxy, $\sigma_i$ is the line-of-sight velocity dispersion, and $F_i$ is the flux in a band (typically chosen to match the wavelength region of extracted kinematics) for the \textit{i}\textsuperscript{th} spaxel or spatial bin. For example, a galaxy with highly ordered stellar motions within a disc has a dominant velocity term relative to the dispersion, and a high overall \vosigma. 
However, due to the surface-brightness weighting, the \vosigma parameter is known to be limited in its ability to discriminate between kinematically distinct galaxy populations, such as galaxies with regular rotation across the galaxy scale radius and those with apparently `decoupled' centrally rotating cores, leading to the development of the proxy for spin parameter, \lamr \citep{emsellem_2007_classification}. This parameter includes a radial weighting and is defined as:  
\begin{equation}
    \lambda_R  = \frac{\sum_{i=1}^{N} F_iR_i|v_i|}{\sum_{i=1}^{N}F_iR_i\sqrt{v_i^2+\sigma_i^2}},
	\label{eq:lamr}
\end{equation}
where $R_i$ is the galactocentric radius for the \textit{i}\textsuperscript{th} spaxel or spatial bin. There are different ways to define this radius; in this work $R_i$ represents the semi-major axis of the ellipse on which the \textit{i}\textsuperscript{th} spaxel lies, which follows galaxy light profiles more naturally than projected circular apertures \citep{cortese_2016_SAMI,Jesse_2017_revisiting}. This \lamr metric was used to quantify galaxy kinematic classes in the pioneering IFS survey $\mathrm{ATLAS}^{\mathrm{3D}}$ \citep{cappellari_2011_ATLAS}, the first such survey of a statistical sample of resolved galaxies, and found that galaxies fall largely into two kinematic categories; `slow rotators' and `fast rotators' \citep{emsellem_2011_census}. This classification scheme uses both the observed \lamr value and projected ellipticity ($\epsilon$) of the galaxy. 
Fast rotators span a wide range of ellipticities but have clear structure in their velocity fields, representing disc-like orbits, and have \lamr values between $\sim 0.1$ and $\sim 1$. Slow rotators, in contrast, are rounder objects with no global, regular rotation signature and therefore have \lamr values close to zero. Slow rotators are also comparatively rarer, forming just $\sim 10$ per cent of the galaxy population in the local Universe \citep{brough_2017_SAMI,jesse_2017_revising}, with the most massive galaxies more likely to be slow rotators \citep{emsellem_2011_census,brough_2017_SAMI,veale_2017_massive}.

Hydrodynamical, cosmological simulations can probe the formation pathways for the slow and fast rotator galaxy populations, and propose mechanisms by which they change across time. \citet{Schulze_2018_kinematics} found that the slow rotator fraction increased from 8 percent at $z = 2$ to 30 per cent at $z = 0$ for early-type galaxies, using a sample of galaxies from the Magneticum Pathfinder simulations \citep{dolag_2015_magneticum}. The accumulation of slow rotators across cosmic time suggests that the mechanism that drives the transformation from fast to slow rotator happens more frequently at later times. One such driver could be successive gas-poor mergers, as suggested by \citet{Naab_2014_analysis}, who linked the observable kinematic features of a galaxy to their evolutionary pathway with a small sample of galaxies from a cosmological `zoom' simulation. Results from `Evolution and Assembly of GaLaxies and their Environments' (EAGLE) simulations \citep{crain_2015_EAGLE,schaye_2015_EAGLE}  also indicate that low spin galaxies generally experienced gas-poor mergers, whereas galaxies at $z = 0$ with normal spin experienced either no mergers or gas-rich mergers. Major mergers in either category of gas-poor or gas-rich exaggerated these effects, with aligned gas-rich major mergers effectively spinning up galaxies \citep{lagos_2018_spin}. An observational study of the orbital distribution of galaxies using Schwarzschild dynamical modelling by \citet{santucci_2022_internal} found that slow rotators assemble at early times  and have evolutionary pathways dominated by gas-poor mergers, whereas fast rotators tend to grow via gas accretion. Overall, the evolution of \lamr within a galaxy population over cosmic time reflects the combination of the above processes, and can therefore be used as a probe of galaxy assembly.

Testing our understanding of the evolution of spin in galaxies is observationally challenging, as it requires spatially resolved spectra of the stellar continuum at increasingly higher redshifts. Here we present measurements of \lamr and \vosigma from the `Middle Ages Galaxy Properties with Integral field spectroscopy' (MAGPI) survey\footnote{Based on observations obtained at the Very Large Telescope (VLT) of the European Southern Observatory (ESO), Paranal, Chile (ESO program ID 1104.B-0536)} \citep{MAGPI}, which targets massive galaxies at a lookback time of 3-4 billion years, with resolution comparable to redshift zero surveys. By carefully comparing MAGPI galaxies to the local Universe `Mapping Nearby Galaxies at Apache Point Observatory' (MaNGA) sample \citep{bundy_2015_manga}, we  test whether there has been a systematic change in the kinematic states of galaxies from the Universe's middle ages to now. This is an initial look at the evolution of \lamr with MAGPI, using the first $\sim 60$ per cent of objects available from the survey (34/56 fields). We leave explorations of environment and detailed comparisons to hydrodynamical, cosmological simulations to a future work. 

In Section \ref{sec:data} we introduce the three data sets used to understand the evolution of \lamr across cosmic time; the MAGPI sample at $z \sim 0.3$, the MaNGA sample at $z \sim 0$, and the EAGLE simulation data used to connect them. In Section \ref{sec:methods} we present our methods for measuring \lamr, sample selection, correcting for seeing effects, and how we make the comparison between MAGPI and MaNGA to account for progenitor bias. We discuss our results in Section \ref{sec:results} and present our conclusions in Section \ref{sec:conclusions}. To convert between angular and physical scales for MAGPI data in this work, we adopt a flat $\Lambda$CDM cosmology with $H_0 = 70\ \kms \mathrm{Mpc}^{-1}$ and $\Omega_m = 0.3$. 

\section{Data}
\label{sec:data}
To understand the possible evolution in the proxy for spin parameter \lamr we compared a sample of MAGPI galaxies to a sample chosen from the MaNGA survey. To connect the two observational data sets across 4 Gyr of cosmic history we used the EAGLE simulations. We introduce each data set below. 

\subsection{MAGPI}
\label{sec:data:MAGPI}
 The MAGPI survey\footnote{\url{https://magpisurvey.org}} is a VLT/MUSE Large Program which targets  galaxies at $z \sim 0.3$. The 60 primary targets have stellar masses estimated at $\mathrm{M}_{\star}>7 \times 10 ^{10} \Msun$ and correspond to galaxies with spatially resolved stellar continuum spectroscopy. In addition to those, there are hundreds of foreground and background sources. The survey uses MUSE in the wide-field mode with spectra in the range $4650 - 9300\,$\AA\ in steps of $1.25\,$\AA\ per pixel, and ground-layer adaptive optics. We measure the PSF field by field; an indicative point-spread function (PSF) for the survey is $\sim 0.6$ arcsec, which is equivalent to $4.5$ kpc at $z \sim 0.3$ with the assumed cosmology. Each field is observed across six observing blocks of $2 \times 1230$ s exposures with rotational dithers, with a total on-source integration time of 4.4 hours per field.

We provide a brief overview of the data reduction process, with full details to be given in Mendel et. al. (in prep.). Reduction is based on the MUSE reduction pipeline \citep{weilbacher_2020_data}, with sky-subtraction completed using Zurich Atmosphere Purge sky-subtraction software \citep{soto_2016_zap}. Segmentation maps and estimates of galaxy structural parameters are created using {\sc{ProFound}}\footnote{\url{https://github.com/asgr/ProFound}} \citep{robotham_2018_profound}. Each source identified within a field is post-processed to be situated at the centre of a `minicube' extracted from the main MUSE cube, which is sized to encompass the maximum extent of the dilated segmentation map determined by {\sc{ProFound}}. This work uses data from spatially resolved primaries and secondaries within the 34 available MAGPI fields as of October 2023 with reduced data and team available data products. The cuts we use on these data are described in Section~\ref{sec:methods:lamr}.

\subsection{MaNGA}
\label{sec:data:MaNGA}

We use the \lamr and \vosigma catalogue of \citet{Fraser_2023_manga} as the local Universe comparison sample, measured from the MaNGA survey. We chose MaNGA as the comparison sample because it contains a large number of high mass systems, allowing us to create random subsets of the sample that represent potential evolutionary endpoints of MAGPI galaxies, and has comparable physical resolution  to the MAGPI survey (see \citealt{MAGPI} figure 2). 

The MaNGA survey observed over 10000 galaxies between redshifts $0.01 < z < 0.15$, with spectral coverage of $3600 - 10300\,$\AA\. The primary sample generally extends of $\sim 1.5$ half-light radii \citep{Yan_2016,Wake_2017}. Reduced and derived properties are available from the MaNGA Data Reduction Pipeline and MaNGA Data Analysis Pipeline, respectively \citep{Law_2015,Westfall_2019}. The average seeing conditions for the survey in the $r$-band give a PSF FWHM of $\sim 2.5$ arcsec. 

Details on how the MaNGA values were measured can be found in \citet{Frazser-Mckelvie_2022_beyond}. As with the MAGPI method described in Section \ref{sec:methods:lamr}, \textit{r}-band measurements of \lamr and \vosigma (defined in Eq.~\ref{eq:vsigma} and Eq.~\ref{eq:lamr})  were made by considering spaxels within the half-light ellipse, and corrected for seeing using 
the kinematic corrections code published by \citet{harborne_2020_recovering}\footnote{\url{https://github.com/kateharborne/kinematic_corrections}}. We took elliptical Petrosian half-light radii for MaNGA objects from the survey's data analysis pipeline summary table, \texttt{drpall\_v3\_1\_1}.\footnote{\url{https://www.sdss4.org/dr17/manga/manga-data/catalogs/}} The MaNGA sample from the study by  \citet{Frazser-Mckelvie_2022_beyond} includes 4741 objects at a median redshift of 0.04.\footnote{Catalogue values were accessed from \citet{Fraser_2023_manga}.}

\subsection{EAGLE}
\label{sec:data:EAGLE}

We use the EAGLE simulations  to link our observed samples of galaxies by tracing the evolution of MAGPI-like EAGLE galaxies to the local Universe, described in Section \ref{sec:methods:prog_bias}. The purpose of using the EAGLE simulations is to account for the varied evolutionary pathways a galaxy might take between the Universe's middle ages and present day, as opposed to adopting a simple stellar mass growth model.

We choose the EAGLE simulations because the galaxy stellar mass function has been shown to accurately match observations up to and beyond $z = 0.5$ \citep{furlong_2015_evolution}, which comfortably includes the redshift span we consider in this work. We do not expect this choice of simulation to be critical, as there is only minor stellar mass growth between redshift $0.3$ and redshift $0$.

In particular, we use the cosmological hydrodynamic EAGLE reference model Ref-L100N1504 and the catalogue of stellar masses and \lamr values created by \citet{lagos_2018_spin}. This reference model has gas particle mass of $1.8 \times 10^{6}\,\Msun$ and dark matter particle mass of $9.7 \times 10^{6}\,\Msun$, with a volume given by sides of $L =  100\,\mathrm{cMpc}^{3}$.  To trace between the MAGPI sample and MaNGA sample we use snapshot 25 at $z = 0.27$, which is closest to the nominated 0.3 target redshift of the MAGPI survey, and snapshot 28 corresponding $z = 0$ to link to the MaNGA sample. In total, the EAGLE sample comprises of 9409 galaxies traceable between the two redshifts, with stellar masses between $9.08 < \log_{10} (M_{\star}/\Msun) < 12.27$ at $z = 0$.

\section{Methods}
\label{sec:methods}
In this section we describe the methods used on MAGPI objects to construct the 2D stellar kinematics and estimate the structural parameters, such as ellipticity and half-light radius, needed for the \lamr and \vosigma measurements. We then describe how the observed values were corrected for seeing conditions, and the method used to account for progenitor bias when comparing the MAGPI galaxies at 3-4 Gyr lookback time to the local Universe MaNGA sample.

\subsection{Stellar kinematics}
\label{sec:methods:stellar_kinematics}

Full details of the MAGPI kinematic extraction can be found in Mendel et. al. (in prep.), but we give a brief overview here. Stellar kinematics are obtained using a customised version of {\sc{gistPipeline}}\footnote{\url{https://abittner.gitlab.io/thegistpipeline/}} v2.1 \citep{bittner_2019_gist} which uses {\sc{pPXF}}\footnote{\url{https://pypi.org/project/ppxf/}} for full-spectrum fitting \citep{ppxf_1,ppxf_2}. A reduced version of IndoUS is used for the stellar template library \citep{valdes_2005_indous}. Spaxels within the {\sc{ProFound}} derived half-light (R50) radius are co-added to construct an aperture spectrum, the fit of which provides the initial redshift estimate for the object.

A subset of templates for use in the fitting of individual spaxels is then determined using the method described in \citet{Jesse_2017_revisiting}. For each object, a series of elliptical annuli are created, considering only spaxels with a signal-to-noise $ > 3$. These annuli are centered on the galaxy and their ellipticity is determined from the {\sc{ProFound}} axis ratios. The annuli are grown by half a pixel until either the target signal-to-noise of $25$ is reached, or there are no more valid spaxels to merge. The spectrum of each annular bin is fit and the non-zero weighted templates used in the fit of each annulus are stored. 

Each individual spaxel is then fit using as input a reduced set of stellar templates, consisting of the templates used to fit the annuli to which the spaxel belongs, and the immediately adjacent annulii. Errors for each spaxel are determined using Monte Carlo simulations, whereby realisations of the input spectra were created by adding random noise based on the residuals of the fit.

\subsection{Estimating structural parameters}
\label{sec:methods:structural_parameters}

Using \textit{r}-band images created from the collapsed MUSE spectral cubes, we model the galaxy light with the Python package {\sc{mgefit}},\footnote{\url{https://pypi.org/project/mgefit/}} which parameterises the light as a series of 2D Gaussians \citep{emsellem_1994,MGEpy}; we refer to this model as a `Multi-Gaussian Expansion' (MGE) in the text. Stacked point-sources in each MAGPI field are used to estimate the PSF (modelled with a circular MGE model) so that the resulting galaxy light models are PSF-deconvolved. From the optimal MGE fit we obtain the ellipse containing half the galaxy light, measuring its ellipticity ($ \epsilon_\mathrm{e} = 1 - b/a$, with $a$ and $b$ the semi-major and semi-minor axes, respectively) and semi-major axis, $R_{\mathrm{e}}$. We use this half-light radius definition in all subsequent MAGPI analysis. We note the MGE method we use here is the same as that presented in \citet{derkenne_2023_impact}, which includes further details on the multi-Gaussian fitting process. This way of estimating the half-light radius is slightly different from the Petrosian half-light radius for the MaNGA sample, but we do not expect this difference to have a significant impact on our results.

Stellar masses for MAGPI objects are calculated using the  spectral energy distribution fitting software {\sc{prospect}}\footnote{\url{https://github.com/asgr/ProSpect}} with a \cite{chabrier_2002_galactic} initial mass function. The spectral energy distribution is measured using `Galaxy and Mass Assembly' (GAMA) survey imaging in the \textit{ugriZYJHKs} bands, pixel-matched to MAGPI data in order to exactly correspond to MAGPI survey objects \citep{bellstedt_2020_gama}. S\'{e}rsic indices are determined using {\sc{galfit}}\footnote{\url{https://users.obs.carnegiescience.edu/peng/work/galfit/galfit.html}} \citep{Peng_2002_galfit} on the collapsed spectral cube images, modelling both the central galaxy and any neighbouring galaxies within five magnitudes. A single S\'{e}rsic component is used, with a local background subtracted prior to fitting.

\label{sec:methods:parameters}
\subsection{Measuring \texorpdfstring{\lamr}{lambdaR} and \texorpdfstring{\vosigma}{vsigma}}

We use Eq.~\ref{eq:vsigma} and Eq.~\ref{eq:lamr} to measure \vosigma and \lamr from the MAGPI data. In this work we choose the half-light ellipse as our aperture, with spaxel flux measured in the \textit{r}-band. We denote \lamr  and \vosigma measured within the half-light ellipse as \lamre and \vosigmae, respectively. Valid spaxels are defined as having finite velocity and velocity dispersion values, with an uncertainty requirement on the velocity dispersion given by:
\begin{equation}
    \Delta \sigma < 25\,\,\mathrm{kms}^{-1}\, + 0.1 \sigma,
    \label{eq:sigma_quality}
\end{equation}
which follows the data quality requirements used by \citet{Jesse_2017_revisiting}. 

\begin{figure}
	\includegraphics[width=\columnwidth]{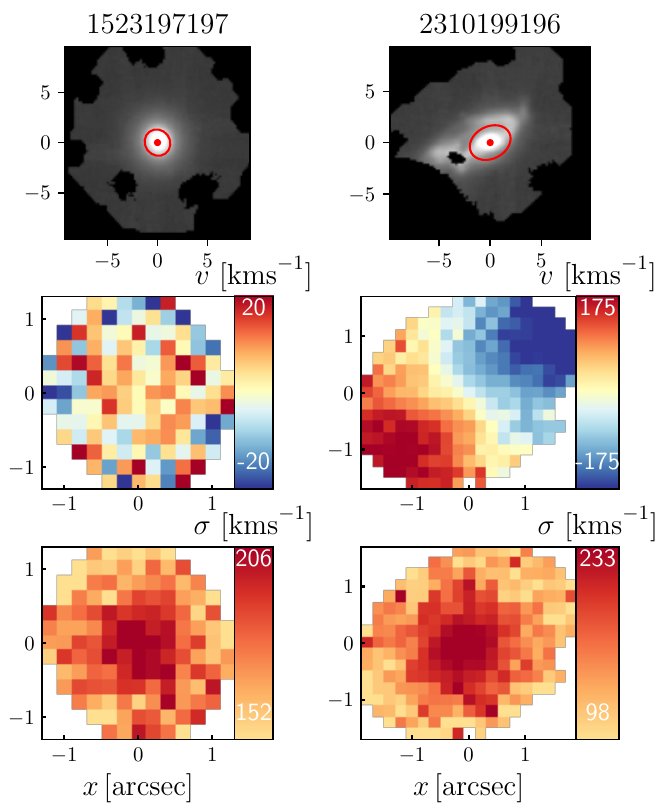}
    \caption{Left column: The top row shows the flux for MAGPI object 1523197197 with the half-light ellipse marked in red. The second row shows its stellar kinematic line-of-sight velocity field, which shows no clear rotation. The third row shows the stellar kinematic velocity dispersion field. Only spaxels within the half-light ellipse are shown. Right column: The same as the left column, but for MAGPI object 2310199196, which has a clear rotation signature in its line-of-sight velocity field.}
    \label{fig:example_fields}
\end{figure}

Uncertainties on the \lamre and \vosigmae values are estimated by creating realisations of the line-of-sight velocity and dispersion values based on their measured uncertainties from the full-spectrum fitting process, described in Section \ref{sec:methods:stellar_kinematics}, and repeating the measurements 2000 times in a Monte-Carlo approach. The uncertainty is given as the standard deviation of the resulting distribution of measurement, divided by $\sqrt{2}$, since this process adds more noise to data that already has noise. As both \lamre and \vosigmae are flux weighted quantities, typical uncertainties are small, with values of $\sim  0.02$. We show two examples of the stellar kinematic fields used to make measurements of \lamre and \vosigmae in Figure \ref{fig:example_fields}, with MAGPI object 1523197197 representing a typical slow rotator and MAGPI object 2310199196 representing a typical fast rotator. 

\begin{figure}
	\includegraphics[width=\columnwidth]{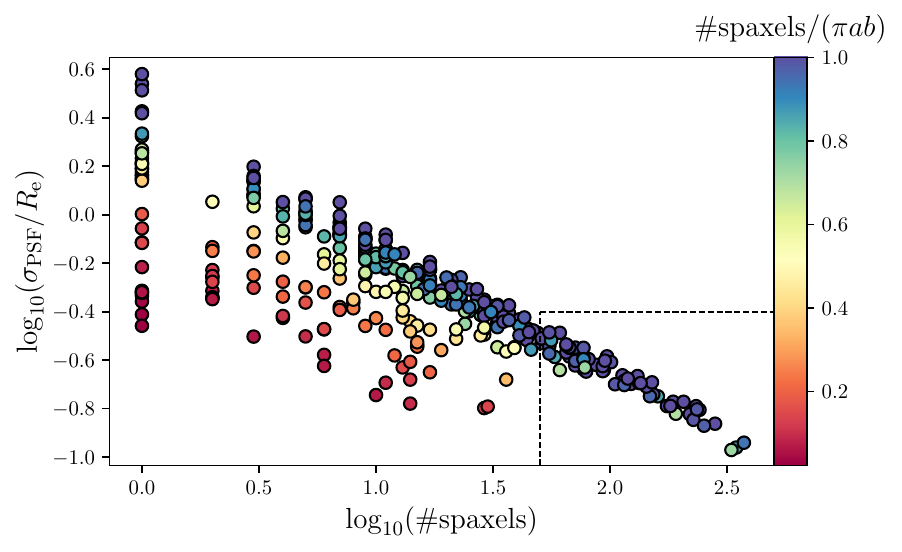}
    \caption{The ($\log_{10}$) number of spaxels within the half-light ellipse against the ($\log_{10}$) resolution, cast as PSF sigma over half-light radius, of all MAGPI objects with a valid \lamre and \vosigmae measurement, coloured by fill-fraction (number of spaxels/area of half-light ellipse). These parameters primarily determine which galaxies can be accurately corrected from seeing-blurred \lamre and \vosigmae values to intrinsic values \citep{harborne_2020_recovering}. The dashed lines show the cuts used in this work, for all galaxies with a fill fraction above 85 per cent.}
    \label{fig:chosen_sample}
\end{figure}

Due to the redshift of the MAGPI survey targets, correcting the measured \lamre and \vosigmae values for the effects of a point-spread function is important, as seeing conditions can artificially reduce observed values. We use the kinematic corrections code of \citet{harborne_2020_recovering} to correct our observed values to their intrinsic ones, based on how resolved each galaxy is, its  S\'{e}rsic index, and ellipticity. We first selected all MAGPI objects with stellar masses $\log_{10} (M_{\star}/\Msun) > 10$. To ensure accurate seeing corrections we require that all galaxies have at least 50 spaxels within the half-light ellipse, with a fill fraction (total spaxel area divided by half-light ellipse area) of more than 85 per cent. We also require that the PSF dispersion (PSF FWHM/2.355) is less than 40 per cent of the semi-major axis of the half-light ellipse, and that the S\'{e}rsic index of each object is less than 8. as  All candidate galaxies with \lamre and \vosigmae measurements are shown in Figure \ref{fig:chosen_sample}. The final sample of MAGPI galaxies used for all subsequent analyses includes \nsample objects. 

We stress that the measurements of \lamre and \vosigmae for MAGPI objects follows closely the methodology of \citet{Frazser-Mckelvie_2022_beyond}, allowing for a fair comparison between the MAGPI and MaNGA observational samples (once progenitor bias and sample selection effects are accounted for - see Section~\ref{sec:methods:prog_bias}). However, due to the redshift of the MAGPI sample, the rest-frame kinematics of the MAGPI sample are in a different band to those of MaNGA, and hence probe subtly different stellar populations. We expect the effect of this difference on the comparison of their \lamre distributions to be small.

\label{sec:methods:lamr}
\subsection{Accounting for progenitor bias}
\label{sec:methods:prog_bias}

To measure the degree of evolution between galaxies in the middle ages of the Universe and today it is important to take into account progenitor bias \citep{dokkum_1996_fundamental}.  The MAGPI sample consists of massive galaxies at a lookback time of 3-4 Gyr, and so choosing a mass-matched sample of massive galaxies in the local Universe disregards any possible evolution in morphology, age, mass, and size that might have occurred across the intervening time. Furthermore, galaxy evolution is inherently stochastic, in the sense it is not deterministic that a particular type of galaxy at one epoch will exactly evolve into another type at later times. This problem has been tackled using hydrodynamical cosmological simulations and tracing the number density of galaxies through cosmic time \citep{torrey_2015_prog_bias,wellons_2017_prog_bias}. To account for the complexities of comparing galaxy populations across time we use a sampling method developed by Khalid et. al. (in prep.), who found that the major driver of progenitor bias is stellar mass. By accounting for evolution in stellar mass alone, through use of cosmological simulations, the evolution of parameters such as \lamr can be accurately inferred. This result is demonstrated in the above work. 

Our implementation of this method is as follows. First, we estimate the probability density function (PDF) for stellar mass of the MAGPI galaxy sample. The PDF is estimated using Gaussian kernels and Silverman's rule for the bandwidth calculation, implemented through {\sc{scipy}} \citep{2020SciPy-NMeth}. We also estimate the stellar mass PDF of the entire EAGLE sample at the redshift snapshot most similar to the MAGPI target redshift ($ z = 0.27$). We then draw a random sample of EAGLE galaxies weighted by the ratio of the MAGPI PDF and EAGLE PDF, with the weighting factor $w_1$ given by:

\begin{equation}
    w_1 = \frac{\mathrm{PDF}(M_{\star,\mathrm{MAGPI}})}{\mathrm{PDF}(M_{\star,\mathrm{EAGLE,z=0.27}})}.
    \label{eq:weighting_1}
\end{equation}

Once normalised so that the probabilities sum to one, the weighting factor on the randomly drawn EAGLE galaxies forces the sample to have the same statistical properties as the MAGPI one in terms of stellar mass. While the drawn EAGLE sample at this point can theoretically have almost arbitrary size, we match it to the MAGPI sample size to build in the effects of having a small number of objects on the end results.

EAGLE galaxies can be traced across snapshots by using a unique combination of group and sub-group IDs, and snapshot number. In this way we let the EAGLE sample evolve to redshift zero and re-measure the stellar mass PDF in the redshift zero snapshot. We again use  weights ($w_2)$ to randomly draw MaNGA galaxies and force the drawn sample to have the same underlying stellar mass statistics as the evolved EAGLE sample:

\begin{equation}
    w_2 = \frac{\mathrm{PDF}(M_{\star,\mathrm{EAGLE\,sample,z=0}})}{\mathrm{PDF}(M_{\star,\mathrm{MaNGA}})}.
    \label{eq:weighting_2}
\end{equation}

\begin{figure}
	\includegraphics[width=\columnwidth]{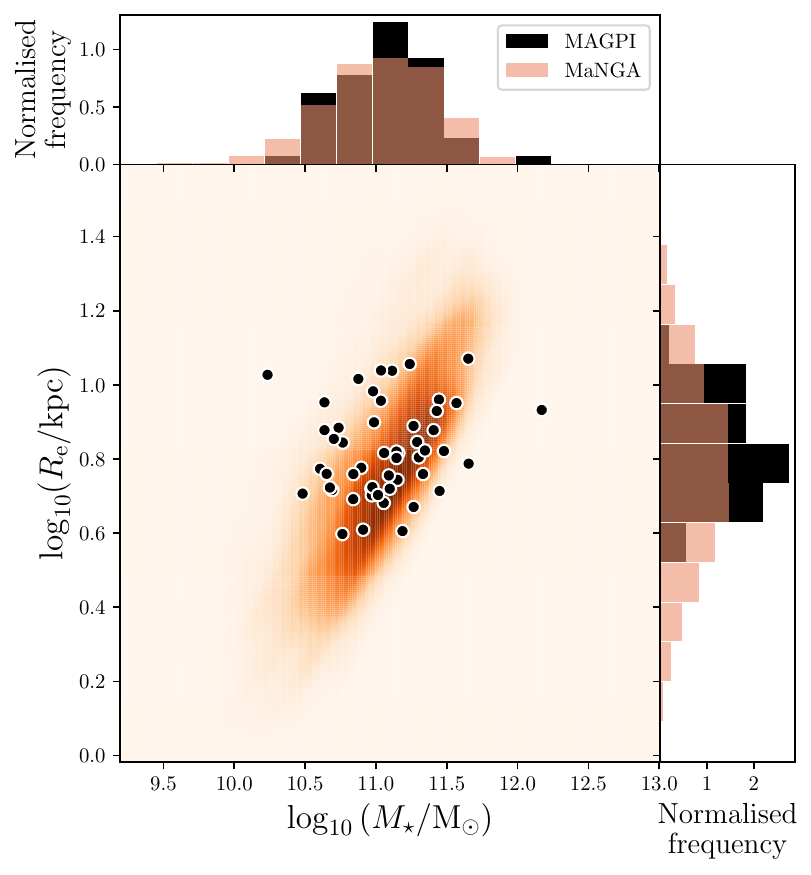}
    \caption{The half-light radius, defined as the semi-major axis of ellipse containing half the galaxy light, against the stellar mass of the MAGPI and MaNGA samples used in this work. All 100 random draws from the MaNGA sample are shown as the background 2D kernel density estimate in orange. The MAGPI sample has a particularly massive object that has no such massive counterpart in the MaNGA sample.}
    \label{fig:mass_size_plane}
\end{figure}

The drawn MaNGA sample represents a possible evolutionary pathway of the MAGPI sample as informed by the EAGLE simulations. However, as noted above, the evolutionary pathways of galaxies are not deterministic, and so a single measurement can only represent one outcome among many. To statistically explore the possible evolutionary scenarios, we repeat the sampling method (randomly re-selecting the EAGLE galaxies to trace, and the local sample of MaNGA galaxies) 100 times to understand the distribution of outcomes, which we present in Section \ref{sec:results}. No cuts are made on the sample in terms of age, star formation rate, or morphology, so that each drawn sample contains a mixture of galaxies. We show the stellar mass and half-light size distribution of the MAGPI galaxies and all drawn MaNGA galaxies in Figure \ref{fig:mass_size_plane}.  

We should note that we have only accounted for stellar mass when choosing a MaNGA sample designed to avoid progenitor bias. Tracing by stellar mass alone only works if the MAGPI sample is representative in its physical parameters compared to the EAGLE one at the same epoch. One key point worth further consideration is that the amount of mass growth an EAGLE galaxy experiences is proportional to its star formation rate, as shown in figure 11 of \citet{crooom_2021_SAMI}, and galaxy kinematics can also depend on star formation rate \cite{wang_2020_kinematic_morphology}. Star-forming galaxies tend to undergo a greater mass growth than passive galaxies  for $z < 1$, and so if the MAGPI sample is over-represented in star-forming objects compared to the EAGLE simulations, we might expect to be slightly biased in the traced mass growth.

To explore this issue,  we characterise the MAGPI sample by using offsets (i.e. distance) from the star-forming main sequence as measured from spectral diagnostics in the MAGPI MUSE data \citep{mun_2024_stars}. The classifications are defined using the following intervals in star formation rate (SFR) from \citet{mun_2024_stars}:
\begin{itemize}
    \item Star-burst: $\Delta$SFR > 0.5 dex
    \item Main-sequence: -0.5 dex < $\Delta$SFR < 0.5 dex
    \item Green-valley: -1.1 dex < $\Delta$SFR < -0.5 dex
    \item Passive: $\Delta$ SFR < -1.1 dex
\end{itemize}
For the sample used in this work, we find 21 objects that are either star-burst, main-sequence, or `green-valley'  objects, and 30 passive objects.
This breakdown is similar to the 40-60 per cent of passive galaxies at the relevant mass range in the EAGLE simulations for {$z \sim 0.1 - 1$}  \citep{furlong_2015_evolution}. We conclude the MAGPI sample is not over-represented in star-forming objects with respect to the EAGLE simulations for stellar masses $\log_{10}M_{\star}/\Msun > 10$. A more complete treatment of the nuanced relationships between stellar mass, star formation rate, and \lamr might be possible with the full MAGPI sample. 

\section{Results and Discussion}
\label{sec:results}

\begin{figure}
	\includegraphics[width=\columnwidth]{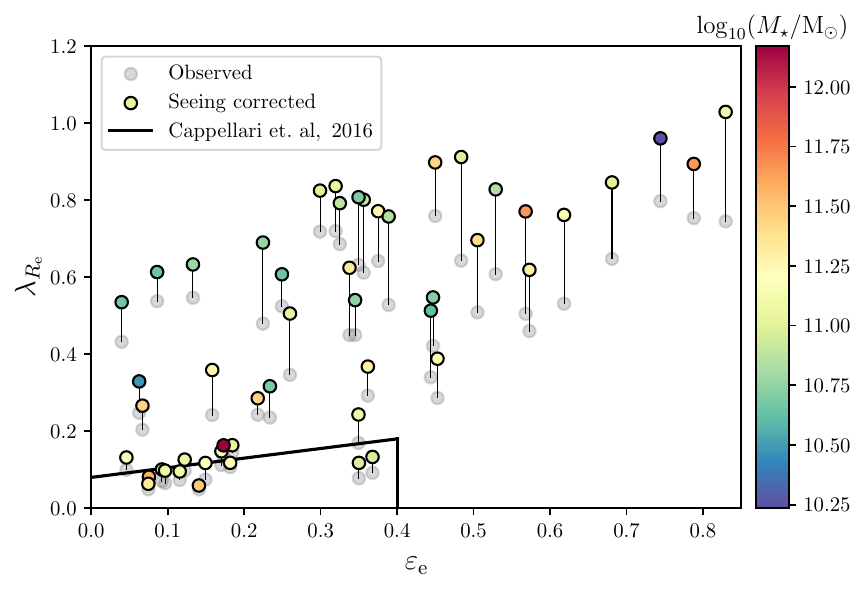}
    \caption{The proxy for spin parameter, \lamre, against ellipticity, \epsilone, both measured within the half-light ellipse,for the MAGPI sample. Objects in the lower left corner are round and pressure supported. Objects in the upper right corner are highly flattened with ordered stellar motions. The observed values are shown in grey, with the seeing-corrected values shown coloured by their stellar mass. The cut between the fast rotating and rarer slow rotating population is shown in black, following \citet{cappellar_2016_structure}. The cut definition is given in Eq.~\ref{eq:slow_rotator_cut}. }
    \label{fig:Lamr_ellip_distribution}
\end{figure}

We show the \lamre distribution against galaxy half-light ellipticity for the MAGPI galaxies in Figure \ref{fig:Lamr_ellip_distribution}. The seeing-corrected values are always higher than the uncorrected values, with smaller corrections typical for lower \lamre values. We find a few objects at the extreme end of the distribution, with \lamre $> 0.8$ and ellipticity greater than $\sim 0.6$, which represent fast rotating, highly flattened systems, and typically tend to be Sb or Sc in their Hubble type \citep{falcon_2019_CALIFA}. One object has a notably high corrected \lamre value of $\sim 1$, which implies negligible dispersion compared to stellar rotation, and could be due to uncertainties on the correction process itself \citep{harborne_2020_recovering}. 

We use the cuts proposed by \citet{cappellar_2016_structure} to discriminate between fast rotators and the rarer population of slow rotating galaxies, as we found this boundary to well separate the slow and fast rotators for the MAGPI sample, based on visual inspection of the velocity fields. However, we note that for the `Sydney-AAO Multi-object Integral field spectrograph survey' \citep[SAMI;][]{croom_2012_SAMI}, a slightly larger region was identified as being optimal when classifying slow rotators \citep{jesse_2021_statistical}. Given this difference between surveys there is not a strict universal region inhabited by purely slow rotating galaxies, and the choice of region should be driven by the actual kinematics of the galaxies in the sample. The seeing correction may also over-correct slow-rotating galaxies, so the definition chosen has subtle impact on the fraction of slow rotators found \citep{jesse_2021_statistical}. Using the \citet{cappellar_2016_structure} metric, a slow rotator has:
\begin{equation}
    \lamre < 0.08 + \epsilone/4\,\,\mathrm{with}\,\,\epsilone < 0.4.
	\label{eq:slow_rotator_cut}
\end{equation}

\begin{figure}
	\includegraphics[width=\columnwidth]{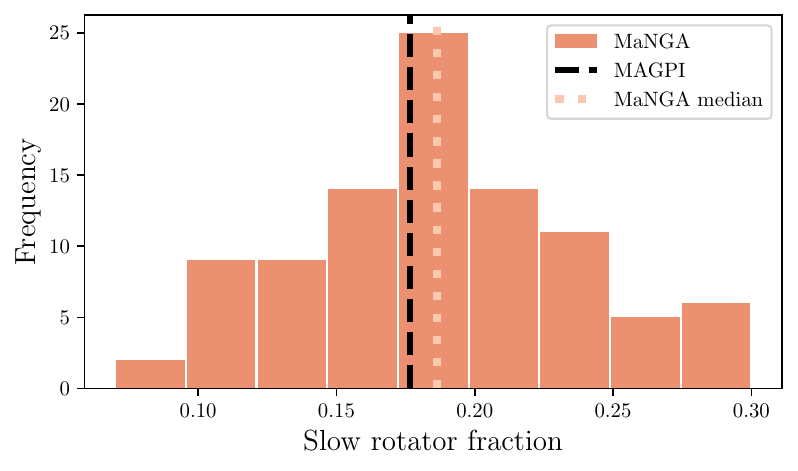}
    \caption{The slow rotator fraction of the MAGPI sample is shown as a black dashed line. All slow rotator fractions from the 100 drawn MaNGA samples are shown as a histogram with median of \mangasrfrac per cent and standard deviation of \mangasrfracstd per cent. The median is marked by the orange dotted line.}
    \label{fig:slow_rotator_fraction}
\end{figure}

In the MAGPI sample we find a slow rotator fraction of $\srfrac^{+6.6}_{-4.1}$ per cent (\srfracN/\nsample objects), with uncertainty estimated using the binomial distribution uncertainty estimate technique from \citet{cameron_2011_estimation}. Without seeing corrections applied, we find a slow rotator fraction of $24^{+0.6}_{-10}$ per cent (12/51 objects), and so the applied corrections do not significantly impact our conclusions. We compare the MAGPI fraction to slow rotator fractions derived from the selected MaNGA samples using the same cuts, shown in Figure \ref{fig:slow_rotator_fraction}. The MaNGA sample across 100  draws has a similar slow rotator fraction at \mangasrfrac per cent with a standard deviation of \mangasrfracstd per cent. However, the lowest slow rotator fraction observed in the MaNGA sample is $\sim 5$ per cent, and the largest is $\sim 30$ per cent, which hints at the pitfalls of comparing small samples in a non-probabilistic manner. Given this large range of possible slow rotator fractions, taking any single draw of the MaNGA sample could lead to interpreting a dramatic evolution in the \lamre parameter, whereas the distribution of samples does not support such an evolution.

\begin{figure}
	\includegraphics[width=\columnwidth]{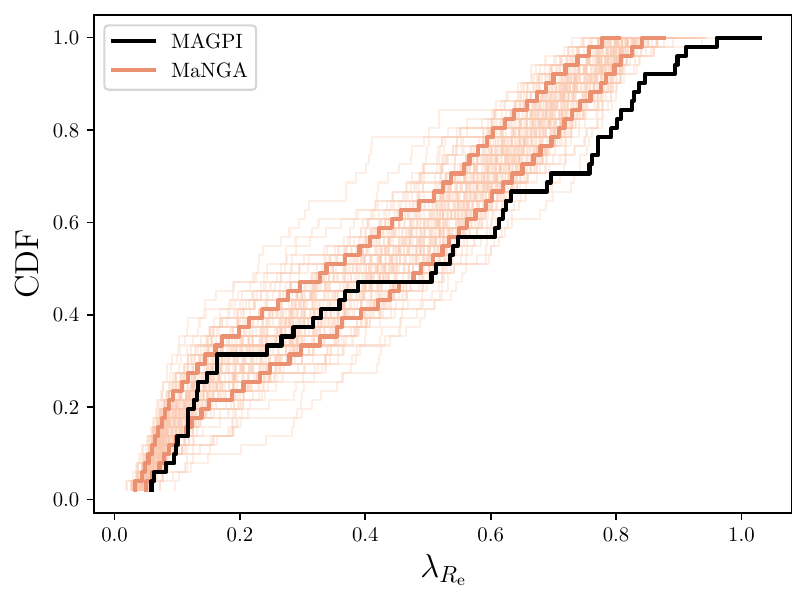}
    \caption{The cumulative distribution function of the MAGPI sample is shown as the black solid line. Each of the 100 random MaNGA draws are shown in light orange, with the 16\textsuperscript{th} and 84\textsuperscript{th} percentiles of the distribution at each \lamr value shown as a bold orange line. For most of the distribution, the MAGPI values fall within the $1\sigma$ region of the MaNGA draws, except for very high spin galaxies.}
    \label{fig:lamR_CDF}
\end{figure}

The entire \lamre distribution can also be compared between the two observational samples, as shown in Figure \ref{fig:lamR_CDF}. The cumulative distribution function (CDF) of the MAGPI sample generally falls within the one-sigma region of the MaNGA drawn samples, except at \lamre values greater than $\sim 0.6$, where the MAGPI sample shows an excess of higher \lamre values compared to the MaNGA sample. This is potentially a hint of the fast rotating population spinning down between redshift 0.3 and the local Universe; however statistically the distributions are not separable. We use an Anderson-Darling test statistic to determine if each of the drawn MaNGA samples and the observed MAGPI sample could be drawn from the same underlying probability distribution, as this statistical test is particularly sensitive to the tails of  distributions \citep{scholz_1987_anderson}. We use an alpha value for significance of 0.01, and across the 100 random draws fail to reject the null hypothesis (that the observed samples are drawn from the same underlying distribution) in \pvalue per cent of cases. 

We then explore only MaNGA and MAGPI \lamre values above 0.5, where the difference is most pronounced, and repeat the statistical tests. Across 100 random draws we fail to reject the null hypothesis in 82 per cent of cases. Although visually there appears to be a significant difference between the distributions at high \lamre, we note that we have the benefit of being able to randomly draw MaNGA samples and thus construct an understanding of the variance of the distributions, whereas we have only one MAGPI sample to compare against. There are also very few galaxies at the very high lambda end where the difference is visually most striking ( 11 with \lamre  $> 0.8$ in MAGPI). These small number statistics also make it difficult to form strong conclusions about this high \lamre trail in MAGPI relative to the MaNGA sample, with statistical tests unlikely to reject the null hypothesis. However, the hint of a difference remains interesting and worthy of future exploration with the complete MAGPI sample.

The high \lamre tail observed in the MAGPI distribution is also partly a function of the kinematic corrections applied to the MAGPI data, which we point out is not without uncertainty. \citet{harborne_2020_recovering} note a residual $1\sigma$ scatter of $\sim 0.02$ dex in \lamre after trends in ellipticity, aperture, and S\'{e}rsic index are accounted for. Realistically these errors are probably larger for the MAGPI sample, as the kinematic corrections code assumes no uncertainty (or very highly resolved) photometric parameters.

\begin{figure}
	\includegraphics[width=\columnwidth]{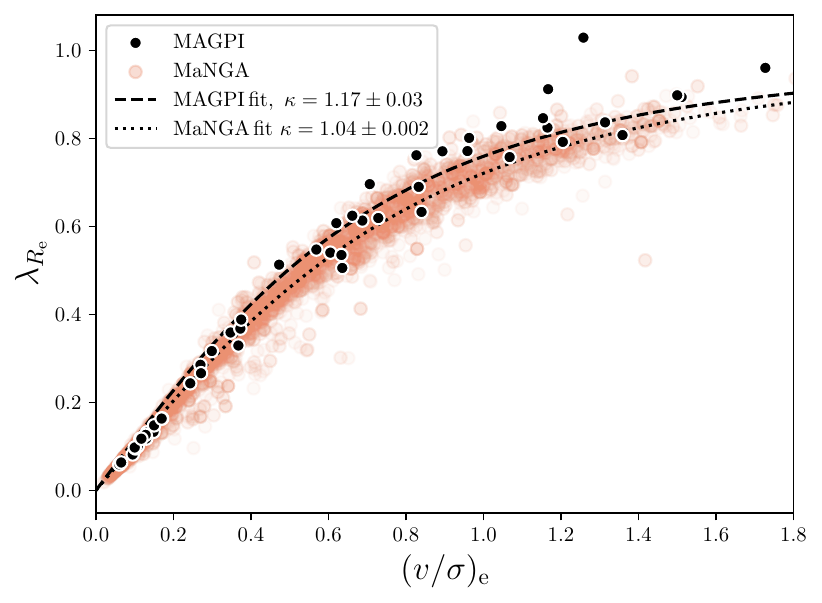}
    \caption{The relationship between \lamre and \vosigmae for MAGPI galaxies in black and MaNGA galaxies in orange. The dashed line shows the best fit of Eq.~\ref{eq:kappa} for MAGPI objects, with the dotted line for all MaNGA objects across the 100 draws.}
    \label{fig:kappa}
\end{figure}

In Figure \ref{fig:kappa} we show the relationship between \lamre and \vosigmae for MAGPI galaxies compared to the randomly drawn MaNGA galaxies. As in \citet{emsellem_2007_kappa} equation B1, we relate the \lamr parameter to \vosigma by:

\begin{equation}
    \lambda_R  = \frac{\langle RV\rangle}{\langle R \sqrt(v^2 + \sigma^2\rangle} \approx \frac{\kappa (v/\sigma)}{\sqrt{1 + \kappa ^2(v/\sigma)^2}}.
	\label{eq:kappa}
\end{equation}

The best-fitting $\kappa$ value for MAGPI is $1.17 \pm  0.03$, compared to the MaNGA value of $1.04 
 \pm 0.002$. As with the \lamre cumulative distribution, we find that there exists a slight separation for the higher spin objects. At the same \vosigmae, MAGPI galaxies tend to have higher \lamre values than MaNGA galaxies for \vosigmae values greater than $\sim 0.6$. 

The agreement of slow rotator fractions between the MAGPI and MaNGA samples, when explicitly accounting for progenitor bias, indicates that the population of slow rotators was already in place in the Universe's middle ages, and has not increased in the past 3-4 Gyr of cosmic time. Given that simulations predict the slow rotator population builds up primarily due to dry mergers \citep{lagos_2018_spin}, our result suggests the mechanism of dry mergers primarily occurred earlier in the Universe's history for massive galaxies. 

We note however that our analysis, by construction, is limited to exploring kinematics within the half-light radius only, and does not preclude changes to the stellar kinematics that may have occurred at larger radial scales. However, obtaining a large sample out to multiple half-light radii at this redshift remains challenging. Furthermore, we have necessarily limited the MAGPI sample to well resolved (less compact) galaxies in order to make accurate measurements of \lamre. Compact galaxies typically have similar \lamre values compared to their less compact counterparts \citep{cappellari_2013_XX}, but their absence from the sample could increase the importance of the massive galaxies in the distribution we study, with massive galaxies more likely to be slow rotators \citep{emsellem_2011_census, veale_2017_massive}. This potential bias is a limitation of studying the kinematics of galaxies beyond the local Universe.

When selecting comparison samples, we have made no consideration of the potentially different environments of MAGPI and MaNGA galaxies. A positive correlation has been noted between the fraction of slow rotators and environmental density in some studies \citep{cappellari_2011_kinematic_density,fogarty_2014}, but it is unclear whether this is actually due to an underlying correlation between stellar \textit{mass} and environment instead \citep{brough_2017_SAMI,Greene_2017}, where as noted above, more massive galaxies are more likely to be slow rotators. Recently, \citet{vaughan_2024} used a generalised linear model with binary classification to understand the relationship between environmental surface density and the kinematic classes of SAMI galaxies. They found that at fixed stellar mass, environmental surface density is not a useful predictor for the kinematic class of a galaxy. Future work on a larger sample will be needed to understand if this is also the case for MAGPI galaxies.

We report that the population of massive slow rotating galaxies was already in place at $z \sim 0.3$, which is tension with the predictions of the Magneticum Pathfinder simulations \citep{dolag_2015_magneticum}. \citet{Schulze_2018_kinematics} in particular predict quite a strong build up of the slow rotator population using Magneticum simulation early-type galaxies. Between redshift 2 and 0, the frequency of slow rotators changes from 8 to 30 per cent, respectively, with an appreciable accumulation of slow rotators between redshifts 0.5 and 0 (see their figure 6).  A slow rotator fraction of 30 per cent at redshift zero is larger than the MaNGA slow rotator fraction in this work of 19 per cent, and of local Universe slow rotator fraction estimates more generally \citep{emsellem_2011_census,Jesse_2017_revisiting,Frazser-Mckelvie_2022_beyond}. Some of the stronger evolution seen by \citet{Schulze_2018_kinematics} may be due to the sample of early-types used. 

\citet{Schulze_2018_kinematics} also note a general trend of fast rotator galaxies redistributing stellar angular momentum across cosmic time, in that the entire population shifts towards lower \lamr values as redshift approaches zero. In the Magneticum Pathfinder simulations this shift happens both gradually, across the entire span from $z = 2$ to $z = 0$, and suddenly, whereby a galaxy transitions from a fast rotator to a slow rotator in less than 0.5 Gyr under the action of a major merger \citep{Schulze_2018_kinematics}. 

In a detailed comparison of the slow rotator fractions from different hydrodynamical cosmological simulations, \citet{jesse_2018_cosmological} found that in general the simulation slow rotator fractions do not match the observed distributions well. This could be due to the many effects that impact how an observed  \lamr measurement is made compared to a \lamr measurement on a simulated galaxy. Some of these effects can be alleviated through software like {\sc{simspin}}\footnote{\url{https://github.com/kateharborne/SimSpin}} \citep{harborne_simspin}, which can generate mock IFS data from cosmological simulations in a consistent manner. 

If the relatively higher MAGPI \lamre values compared to MaNGA are indicative of a subtle difference in their intrinsic distributions, it is interesting to speculate that the higher \lamre objects might then be more sensitive to whatever mechanism causes the change in how stellar angular momentum is distributed across time. As mentioned above, high \lamre objects tend to be spirals in morphological type \citep[e.g.][]{falcon_2019_CALIFA}, and so this could be an indication of disc fading \citep{crooom_2021_SAMI} or an uneven effect of the efficiency of dry mergers on galaxy spin. Speculation aside, we stress again that  although there is an excess of high \lamre MAGPI objects compared to MaNGA ones, we do not find a statistically significant difference between the sample distributions and therefore conclude that there has not been a significant change in \lamre values across this time. 

An analysis of the excess kurtosis of the stellar velocity distribution profile, $h_4$, by \citet{deugenio_2023_evolution}, found that there is $7\sigma$ evidence of a change in $h_4$ between redshifts 0.8 and 0 for quiescent galaxies with $\log_{10}(M_{\star}/\Msun) > 11$. The result suggests stellar anisotropy has increased gradually over cosmic time, perhaps due to minor mergers depositing ex-situ stars at large radii, with evolution occurring possibly as early as $z = 0.8$. \citet{deugenio_2023_evolution} note their result is slightly in tension with the predictions of the EAGLE simulations, which predict apparent galaxy spin down occurs mainly at $z < 0.5$ \citep{lagos_2018_quantifying}. The results of this work, where we note only a marginal excess of high \lamre values at $z = 0.3$ compared to local Universe values, are consistent with the conclusion of \citet{deugenio_2023_evolution}, in that it suggests evolution of the stellar orbital structure occurred at earlier times. Likewise, \citet{bezanson_2018_spatially} compared the rotational support of quiescent galaxies at $ z \sim 0.8$ and local galaxies, finding that galaxies at $ z \sim 0.8$ had 94 per cent more rotational support at a radius of 5 kpc than their local counterparts, possibly due to a combination of dissipationless merging and the mechanism that ceases star formation. Galaxies at even early times ($z \sim 2$) typically tend to be fast rotators, inferred both from their apparent shapes \citep{van_der_wel_2011_majority}, and also their kinematics \citep{newman_2018_resolving}, implying there must be an accumulation of slow rotators at some point in the intervening time between the early Universe and present day. 

\section{Conclusions}
\label{sec:conclusions}

We compare a sample of \nsample massive MAGPI galaxies at a lookback time of 3-4 Gyr to carefully drawn samples of MaNGA galaxies in the local Universe to determine whether there has been a significant change in the stellar kinematic state of galaxies across this time. We quantify the kinematic state of a galaxy through its \lamre parameter, which is an observational proxy for spin. We use the EAGLE cosmological, hydrodynamical simulations to draw samples of galaxies mass-matched to our MAGPI sample at z=0.3, and trace them through time to $z=0$. This allows us to draw samples of local galaxies from the MaNGA survey to compare the \lamre distribution at $z\sim 0$, accounting for progenitor bias and the diverse and stochastic evolutionary pathways a galaxy can follow across its history. We analyse the MAGPI \lamre distribution against 100 drawn MaNGA \lamre samples.

We assess whether the slow rotator population \citep[using the definition of][]{cappellar_2016_structure} has changed from the middles ages of the Universe to now. We find a slow rotator fraction of 18 per cent  in our MAGPI sample. This is consistent with the median slow rotator fraction from the mass-evolved MaNGA samples, found to be $19 \pm 6$ per cent. Our results indicate there has been no significant evolution of the slow rotator fraction over the past 3-4 Gyr, as uniquely probed by MAGPI. We conclude that the massive ($\log_{10}M_{\star}/\Msun > 10$) slow rotator population of galaxies  was already in place at $z \sim 0.3$, and has not accumulated significantly since then. This result is an important calibration for models of galaxy evolution. 

Finally, we compare the overall distribution of \lamre values between the MAGPI sample and the 100 drawn MaNGA samples. The MAGPI cumulative distribution of \lamre falls within the 1-$\sigma$ expected range of the MaNGA draws for \lamre $ < 0.6$. Above this, there is a slight excess in the MAGPI \lamre values compared to MaNGA ones, which hints that the fastest rotating galaxies may, on average, have spun down in the intervening time. However, the distributions are not statistically separable using an Anderson-Darling test statistic. The possible evolution of \lamre at earlier times will be interesting to explore in future work with the full MAGPI sample for this epoch. The epoch at which the current slow rotator population became established may yet be explored by pushing resolved stellar kinematic studies to high redshifts with the benefit of instruments such as the JWST, and ground-based instruments such as the `MCAO Assisted Visible Imager and Spectrograph' \citep[MAVIS;][]{mcDermid_2020_mavis}.

\section*{Acknowledgements}
We warmly thank the reviewer for their comments which helped improve this work.
We wish to thank the ESO staff, and in particular the staff at Paranal Observatory, for carrying out the MAGPI observations. MAGPI targets were selected from GAMA. GAMA is a joint
European-Australasian project based around a spectroscopic campaign using the Anglo-Australian
Telescope. GAMA was funded by the STFC (UK), the ARC (Australia), the AAO, and the participating
institutions. GAMA photometry is based on observations made with ESO Telescopes at the La Silla
Paranal Observatory under programme ID 179.A-2004, ID 177.A-3016. The MAGPI team
acknowledge support by the Australian Research Council Centre of Excellence for All Sky
Astrophysics in 3 Dimensions (ASTRO 3D), through project number CE170100013. CF is the recipient of an Australian Research Council Future Fellowship (project number FT210100168) funded by the Australian Government. CL, JTM and CF are the recipients of ARC Discovery Project DP210101945. GS acknowledges funding from the Australian Research Council (ARC) Discovery Project DP210101945. AFM acknowledges support from RYC2021-031099-I and PID2021-123313NA-I00 of MICIN/AEI/10.13039/501100011033/FEDER, UE, NextGenerationEU/PRT. FDE acknowledges support by the Science and Technology Facilities Council (STFC), by the ERC Advanced Grant 695671 ``QUENCH'', and by the UKRI Frontier Research grant RISEandFALL. FDE is grateful to C. Conroy for sharing the high-resolution C3K/MIST library. JvdS acknowledges support of an Australian Research Council Discovery Early Career Research Award (project number DE200100461) funded by the Australian Government. Y.P. acknowledges the support from the National Science Foundation of China (NSFC) grant Nos. 12125301, 12192220, 12192222, and the science research grants from the China Manned Space Project with No. CMS-CSST-2021-A07. SMS acknowledges funding from the Australian Research Council (DE220100003). LMV acknowledges support by the German Academic Scholarship Foundation (Studienstiftung des deutschen Volkes) and the Marianne-Plehn-Program of the Elite Network of Bavaria.

%%%%%%%%%%%%%%%%%%%%%%%%%%%%%%%%%%%%%%%%%%%%%%%%%%
\section*{Data Availability}

The MAGPI spectral data underlying this work are available from the ESO Science Archive Facility ({\url{http://archive.eso.org/cms.html}}).

%%%%%%%%%%%%%%%%%%%% REFERENCES %%%%%%%%%%%%%%%%%%
\bibliographystyle{mnras}
\bibliography{lambda_re_evolution} 

\begin{thebibliography}{}
\makeatletter
\relax
\def\mn@urlcharsother{\let\do\@makeother \do\$\do\&\do\#\do\^\do\_\do\%\do\~}
\def\mn@doi{\begingroup\mn@urlcharsother \@ifnextchar [ {\mn@doi@}
  {\mn@doi@[]}}
\def\mn@doi@[#1]#2{\def\@tempa{#1}\ifx\@tempa\@empty \href
  {http://dx.doi.org/#2} {doi:#2}\else \href {http://dx.doi.org/#2} {#1}\fi
  \endgroup}
\def\mn@eprint#1#2{\mn@eprint@#1:#2::\@nil}
\def\mn@eprint@arXiv#1{\href {http://arxiv.org/abs/#1} {{\tt arXiv:#1}}}
\def\mn@eprint@dblp#1{\href {http://dblp.uni-trier.de/rec/bibtex/#1.xml}
  {dblp:#1}}
\def\mn@eprint@#1:#2:#3:#4\@nil{\def\@tempa {#1}\def\@tempb {#2}\def\@tempc
  {#3}\ifx \@tempc \@empty \let \@tempc \@tempb \let \@tempb \@tempa \fi \ifx
  \@tempb \@empty \def\@tempb {arXiv}\fi \@ifundefined
  {mn@eprint@\@tempb}{\@tempb:\@tempc}{\expandafter \expandafter \csname
  mn@eprint@\@tempb\endcsname \expandafter{\@tempc}}}

\bibitem[\protect\citeauthoryear{{Bellstedt} et~al.,}{{Bellstedt}
  et~al.}{2020}]{bellstedt_2020_gama}
{Bellstedt} S.,  et~al., 2020, \mn@doi [\mnras] {10.1093/mnras/staa1466}, \href
  {https://ui.adsabs.harvard.edu/abs/2020MNRAS.496.3235B} {496, 3235}

\bibitem[\protect\citeauthoryear{{Bezanson} et~al.,}{{Bezanson}
  et~al.}{2018}]{bezanson_2018_spatially}
{Bezanson} R.,  et~al., 2018, \mn@doi [\apj] {10.3847/1538-4357/aabc55}, \href
  {https://ui.adsabs.harvard.edu/abs/2018ApJ...858...60B} {858, 60}

\bibitem[\protect\citeauthoryear{{Binney}}{{Binney}}{1978}]{binney_1978_rotation}
{Binney} J.,  1978, \mn@doi [\mnras] {10.1093/mnras/183.3.501}, \href
  {https://ui.adsabs.harvard.edu/abs/1978MNRAS.183..501B} {183, 501}

\bibitem[\protect\citeauthoryear{{Binney}}{{Binney}}{2005}]{binney_2005_rotation}
{Binney} J.,  2005, \mn@doi [\mnras] {10.1111/j.1365-2966.2005.09495.x}, \href
  {https://ui.adsabs.harvard.edu/abs/2005MNRAS.363..937B} {363, 937}

\bibitem[\protect\citeauthoryear{{Bittner} et~al.,}{{Bittner}
  et~al.}{2019}]{bittner_2019_gist}
{Bittner} A.,  et~al., 2019, \mn@doi [\aap] {10.1051/0004-6361/201935829},
  \href {https://ui.adsabs.harvard.edu/abs/2019A&A...628A.117B} {628, A117}

\bibitem[\protect\citeauthoryear{{Brough} et~al.,}{{Brough}
  et~al.}{2017}]{brough_2017_SAMI}
{Brough} S.,  et~al., 2017, \mn@doi [\apj] {10.3847/1538-4357/aa7a11}, \href
  {https://ui.adsabs.harvard.edu/abs/2017ApJ...844...59B} {844, 59}

\bibitem[\protect\citeauthoryear{{Bundy} et~al.,}{{Bundy}
  et~al.}{2015}]{bundy_2015_manga}
{Bundy} K.,  et~al., 2015, \mn@doi [\apj] {10.1088/0004-637X/798/1/7}, \href
  {https://ui.adsabs.harvard.edu/abs/2015ApJ...798....7B} {798, 7}

\bibitem[\protect\citeauthoryear{{Cameron}}{{Cameron}}{2011}]{cameron_2011_estimation}
{Cameron} E.,  2011, \mn@doi [\pasa] {10.1071/AS10046}, \href
  {https://ui.adsabs.harvard.edu/abs/2011PASA...28..128C} {28, 128}

\bibitem[\protect\citeauthoryear{{Cappellari}}{{Cappellari}}{2002}]{MGEpy}
{Cappellari} M.,  2002, \mn@doi [\mnras] {10.1046/j.1365-8711.2002.05412.x},
  \href {https://ui.adsabs.harvard.edu/abs/2002MNRAS.333..400C} {333, 400}

\bibitem[\protect\citeauthoryear{{Cappellari}}{{Cappellari}}{2016}]{cappellar_2016_structure}
{Cappellari} M.,  2016, \mn@doi [\araa] {10.1146/annurev-astro-082214-122432},
  \href {https://ui.adsabs.harvard.edu/abs/2016ARA&A..54..597C} {54, 597}

\bibitem[\protect\citeauthoryear{{Cappellari}}{{Cappellari}}{2017}]{ppxf_2}
{Cappellari} M.,  2017, \mn@doi [\mnras] {10.1093/mnras/stw3020}, \href
  {https://ui.adsabs.harvard.edu/abs/2017MNRAS.466..798C} {466, 798}

\bibitem[\protect\citeauthoryear{{Cappellari} \& {Emsellem}}{{Cappellari} \&
  {Emsellem}}{2004}]{ppxf_1}
{Cappellari} M.,  {Emsellem} E.,  2004, \mn@doi [\pasp] {10.1086/381875}, \href
  {https://ui.adsabs.harvard.edu/abs/2004PASP..116..138C} {116, 138}

\bibitem[\protect\citeauthoryear{{Cappellari} et~al.,}{{Cappellari}
  et~al.}{2011a}]{cappellari_2011_ATLAS}
{Cappellari} M.,  et~al., 2011a, \mn@doi [\mnras]
  {10.1111/j.1365-2966.2010.18174.x}, \href
  {https://ui.adsabs.harvard.edu/abs/2011MNRAS.413..813C} {413, 813}

\bibitem[\protect\citeauthoryear{Cappellari et~al.,}{Cappellari
  et~al.}{2011b}]{cappellari_2011_kinematic_density}
Cappellari M.,  et~al., 2011b, \mn@doi [Monthly Notices of the Royal
  Astronomical Society] {10.1111/j.1365-2966.2011.18600.x}, 416, 1680

\bibitem[\protect\citeauthoryear{{Cappellari} et~al.,}{{Cappellari}
  et~al.}{2013}]{cappellari_2013_XX}
{Cappellari} M.,  et~al., 2013, \mn@doi [\mnras] {10.1093/mnras/stt644}, \href
  {https://ui.adsabs.harvard.edu/abs/2013MNRAS.432.1862C} {432, 1862}

\bibitem[\protect\citeauthoryear{{Chabrier}}{{Chabrier}}{2003}]{chabrier_2002_galactic}
{Chabrier} G.,  2003, \mn@doi [\pasp] {10.1086/376392}, \href
  {https://ui.adsabs.harvard.edu/abs/2003PASP..115..763C} {115, 763}

\bibitem[\protect\citeauthoryear{{Cortese} et~al.,}{{Cortese}
  et~al.}{2016}]{cortese_2016_SAMI}
{Cortese} L.,  et~al., 2016, \mn@doi [\mnras] {10.1093/mnras/stw1891}, \href
  {https://ui.adsabs.harvard.edu/abs/2016MNRAS.463..170C} {463, 170}

\bibitem[\protect\citeauthoryear{{Crain} et~al.,}{{Crain}
  et~al.}{2015}]{crain_2015_EAGLE}
{Crain} R.~A.,  et~al., 2015, \mn@doi [\mnras] {10.1093/mnras/stv725}, \href
  {https://ui.adsabs.harvard.edu/abs/2015MNRAS.450.1937C} {450, 1937}

\bibitem[\protect\citeauthoryear{Croom et~al.,}{Croom
  et~al.}{2012}]{croom_2012_SAMI}
Croom S.~M.,  et~al., 2012, \mn@doi [Monthly Notices of the Royal Astronomical
  Society] {10.1111/j.1365-2966.2011.20365.x}, 421, 872

\bibitem[\protect\citeauthoryear{{Croom} et~al.,}{{Croom}
  et~al.}{2021}]{crooom_2021_SAMI}
{Croom} S.~M.,  et~al., 2021, \mn@doi [\mnras] {10.1093/mnras/stab1494}, \href
  {https://ui.adsabs.harvard.edu/abs/2021MNRAS.505.2247C} {505, 2247}

\bibitem[\protect\citeauthoryear{{D'Eugenio} et~al.,}{{D'Eugenio}
  et~al.}{2023}]{deugenio_2023_evolution}
{D'Eugenio} F.,  et~al., 2023, \mn@doi [\mnras] {10.1093/mnras/stad800}, \href
  {https://ui.adsabs.harvard.edu/abs/2023MNRAS.525.2789D} {525, 2789}

\bibitem[\protect\citeauthoryear{{Derkenne} et~al.,}{{Derkenne}
  et~al.}{2023}]{derkenne_2023_impact}
{Derkenne} C.,  et~al., 2023, \mn@doi [\mnras] {10.1093/mnras/stad1079}, \href
  {https://ui.adsabs.harvard.edu/abs/2023MNRAS.522.3602D} {522, 3602}

\bibitem[\protect\citeauthoryear{{Dolag}}{{Dolag}}{2015}]{dolag_2015_magneticum}
{Dolag} K.,  2015, in IAU General Assembly. p. 2250156

\bibitem[\protect\citeauthoryear{{Emsellem}, {Monnet}  \& {Bacon}}{{Emsellem}
  et~al.}{1994}]{emsellem_1994}
{Emsellem} E.,  {Monnet} G.,   {Bacon} R.,  1994, \aap, \href
  {https://ui.adsabs.harvard.edu/abs/1994A&A...285..723E} {285, 723}

\bibitem[\protect\citeauthoryear{{Emsellem} et~al.,}{{Emsellem}
  et~al.}{2007a}]{emsellem_2007_classification}
{Emsellem} E.,  et~al., 2007a, \mn@doi [\mnras]
  {10.1111/j.1365-2966.2007.11752.x}, \href
  {https://ui.adsabs.harvard.edu/abs/2007MNRAS.379..401E} {379, 401}

\bibitem[\protect\citeauthoryear{{Emsellem} et~al.,}{{Emsellem}
  et~al.}{2007b}]{emsellem_2007_kappa}
{Emsellem} E.,  et~al., 2007b, \mn@doi [\mnras]
  {10.1111/j.1365-2966.2007.11752.x}, \href
  {https://ui.adsabs.harvard.edu/abs/2007MNRAS.379..401E} {379, 401}

\bibitem[\protect\citeauthoryear{{Emsellem} et~al.,}{{Emsellem}
  et~al.}{2011}]{emsellem_2011_census}
{Emsellem} E.,  et~al., 2011, \mn@doi [\mnras]
  {10.1111/j.1365-2966.2011.18496.x}, \href
  {https://ui.adsabs.harvard.edu/abs/2011MNRAS.414..888E} {414, 888}

\bibitem[\protect\citeauthoryear{{Falc{\'o}n-Barroso}
  et~al.,}{{Falc{\'o}n-Barroso} et~al.}{2019}]{falcon_2019_CALIFA}
{Falc{\'o}n-Barroso} J.,  et~al., 2019, \mn@doi [\aap]
  {10.1051/0004-6361/201936413}, \href
  {https://ui.adsabs.harvard.edu/abs/2019A&A...632A..59F} {632, A59}

\bibitem[\protect\citeauthoryear{Fogarty et~al.,}{Fogarty
  et~al.}{2014}]{fogarty_2014}
Fogarty L. M.~R.,  et~al., 2014, \mn@doi [Monthly Notices of the Royal
  Astronomical Society] {10.1093/mnras/stu1165}, 443, 485

\bibitem[\protect\citeauthoryear{{Foster} et~al.,}{{Foster}
  et~al.}{2021}]{MAGPI}
{Foster} C.,  et~al., 2021, \mn@doi [\pasa] {10.1017/pasa.2021.25}, \href
  {https://ui.adsabs.harvard.edu/abs/2021PASA...38...31F} {38, e031}

\bibitem[\protect\citeauthoryear{{Fraser-McKelvie} \&
  {Cortese}}{{Fraser-McKelvie} \&
  {Cortese}}{2022}]{Frazser-Mckelvie_2022_beyond}
{Fraser-McKelvie} A.,  {Cortese} L.,  2022, \mn@doi [\apj]
  {10.3847/1538-4357/ac874d}, \href
  {https://ui.adsabs.harvard.edu/abs/2022ApJ...937..117F} {937, 117}

\bibitem[\protect\citeauthoryear{{Fraser-McKelvie} \&
  {Cortese}}{{Fraser-McKelvie} \& {Cortese}}{2023}]{Fraser_2023_manga}
{Fraser-McKelvie} A.,  {Cortese} L.,  2023, VizieR Online Data Catalog, \href
  {https://ui.adsabs.harvard.edu/abs/2023yCat..19370117F} {p. J/ApJ/937/117}

\bibitem[\protect\citeauthoryear{{Furlong} et~al.,}{{Furlong}
  et~al.}{2015}]{furlong_2015_evolution}
{Furlong} M.,  et~al., 2015, \mn@doi [\mnras] {10.1093/mnras/stv852}, \href
  {https://ui.adsabs.harvard.edu/abs/2015MNRAS.450.4486F} {450, 4486}

\bibitem[\protect\citeauthoryear{Greene et~al.,}{Greene
  et~al.}{2017}]{Greene_2017}
Greene J.~E.,  et~al., 2017, \mn@doi [The Astrophysical Journal Letters]
  {10.3847/2041-8213/aa8ace}, 851, L33

\bibitem[\protect\citeauthoryear{{Harborne}, {van de Sande}, {Cortese},
  {Power}, {Robotham}, {Lagos}  \& {Croom}}{{Harborne}
  et~al.}{2020}]{harborne_2020_recovering}
{Harborne} K.~E.,  {van de Sande} J.,  {Cortese} L.,  {Power} C.,  {Robotham}
  A.~S.~G.,  {Lagos} C.~D.~P.,   {Croom} S.,  2020, \mn@doi [\mnras]
  {10.1093/mnras/staa1847}, \href
  {https://ui.adsabs.harvard.edu/abs/2020MNRAS.497.2018H} {497, 2018}

\bibitem[\protect\citeauthoryear{{Harborne} et~al.,}{{Harborne}
  et~al.}{2023}]{harborne_simspin}
{Harborne} K.~E.,  et~al., 2023, \mn@doi [\pasa] {10.1017/pasa.2023.47}, \href
  {https://ui.adsabs.harvard.edu/abs/2023PASA...40...48H} {40, e048}

\bibitem[\protect\citeauthoryear{{Lagos} et~al.,}{{Lagos}
  et~al.}{2018a}]{lagos_2018_quantifying}
{Lagos} C. d.~P.,  et~al., 2018a, \mn@doi [\mnras] {10.1093/mnras/stx2667},
  \href {https://ui.adsabs.harvard.edu/abs/2018MNRAS.473.4956L} {473, 4956}

\bibitem[\protect\citeauthoryear{{Lagos}, {Schaye}, {Bah{\'e}}, {van de Sande},
  {Kay}, {Barnes}, {Davis}  \& {Dalla Vecchia}}{{Lagos}
  et~al.}{2018b}]{lagos_2018_spin}
{Lagos} C. d.~P.,  {Schaye} J.,  {Bah{\'e}} Y.,  {van de Sande} J.,  {Kay}
  S.~T.,  {Barnes} D.,  {Davis} T.~A.,   {Dalla Vecchia} C.,  2018b, \mn@doi
  [\mnras] {10.1093/mnras/sty489}, \href
  {https://ui.adsabs.harvard.edu/abs/2018MNRAS.476.4327L} {476, 4327}

\bibitem[\protect\citeauthoryear{Law et~al.,}{Law et~al.}{2015}]{Law_2015}
Law D.~R.,  et~al., 2015, \mn@doi [The Astronomical Journal]
  {10.1088/0004-6256/150/1/19}, 150, 19

\bibitem[\protect\citeauthoryear{{McDermid} et~al.,}{{McDermid}
  et~al.}{2020}]{mcDermid_2020_mavis}
{McDermid} R.~M.,  et~al., 2020, \mn@doi [arXiv e-prints]
  {10.48550/arXiv.2009.09242}, \href
  {https://ui.adsabs.harvard.edu/abs/2020arXiv200909242M} {p. arXiv:2009.09242}

\bibitem[\protect\citeauthoryear{{Mun} et~al.,}{{Mun}
  et~al.}{2024}]{mun_2024_stars}
{Mun} M.,  et~al., 2024, \mn@doi [\mnras] {10.1093/mnras/stae1132}, \href
  {https://ui.adsabs.harvard.edu/abs/2024MNRAS.tmp.1153M} {}

\bibitem[\protect\citeauthoryear{{Naab} et~al.,}{{Naab}
  et~al.}{2014}]{Naab_2014_analysis}
{Naab} T.,  et~al., 2014, \mn@doi [\mnras] {10.1093/mnras/stt1919}, \href
  {https://ui.adsabs.harvard.edu/abs/2014MNRAS.444.3357N} {444, 3357}

\bibitem[\protect\citeauthoryear{{Newman}, {Belli}, {Ellis}  \&
  {Patel}}{{Newman} et~al.}{2018}]{newman_2018_resolving}
{Newman} A.~B.,  {Belli} S.,  {Ellis} R.~S.,   {Patel} S.~G.,  2018, \mn@doi
  [\apj] {10.3847/1538-4357/aacd4f}, \href
  {https://ui.adsabs.harvard.edu/abs/2018ApJ...862..126N} {862, 126}

\bibitem[\protect\citeauthoryear{{Peng}, {Ho}, {Impey}  \& {Rix}}{{Peng}
  et~al.}{2002}]{Peng_2002_galfit}
{Peng} C.~Y.,  {Ho} L.~C.,  {Impey} C.~D.,   {Rix} H.-W.,  2002, \mn@doi [\aj]
  {10.1086/340952}, \href
  {https://ui.adsabs.harvard.edu/abs/2002AJ....124..266P} {124, 266}

\bibitem[\protect\citeauthoryear{{Robotham}, {Davies}, {Driver}, {Koushan},
  {Taranu}, {Casura}  \& {Liske}}{{Robotham}
  et~al.}{2018}]{robotham_2018_profound}
{Robotham} A.~S.~G.,  {Davies} L.~J.~M.,  {Driver} S.~P.,  {Koushan} S.,
  {Taranu} D.~S.,  {Casura} S.,   {Liske} J.,  2018, \mn@doi [\mnras]
  {10.1093/mnras/sty440}, \href
  {https://ui.adsabs.harvard.edu/abs/2018MNRAS.476.3137R} {476, 3137}

\bibitem[\protect\citeauthoryear{{Santucci} et~al.,}{{Santucci}
  et~al.}{2022}]{santucci_2022_internal}
{Santucci} G.,  et~al., 2022, \mn@doi [\apj] {10.3847/1538-4357/ac5bd5}, \href
  {https://ui.adsabs.harvard.edu/abs/2022ApJ...930..153S} {930, 153}

\bibitem[\protect\citeauthoryear{{Schaye} et~al.,}{{Schaye}
  et~al.}{2015}]{schaye_2015_EAGLE}
{Schaye} J.,  et~al., 2015, \mn@doi [\mnras] {10.1093/mnras/stu2058}, \href
  {https://ui.adsabs.harvard.edu/abs/2015MNRAS.446..521S} {446, 521}

\bibitem[\protect\citeauthoryear{Scholz \& Stephens}{Scholz \&
  Stephens}{1987}]{scholz_1987_anderson}
Scholz F.~W.,  Stephens M.~A.,  1987, Journal of the American Statistical
  Association, 82, 918

\bibitem[\protect\citeauthoryear{{Schulze}, {Remus}, {Dolag}, {Burkert},
  {Emsellem}  \& {van de Ven}}{{Schulze}
  et~al.}{2018}]{Schulze_2018_kinematics}
{Schulze} F.,  {Remus} R.-S.,  {Dolag} K.,  {Burkert} A.,  {Emsellem} E.,
  {van de Ven} G.,  2018, \mn@doi [\mnras] {10.1093/mnras/sty2090}, \href
  {https://ui.adsabs.harvard.edu/abs/2018MNRAS.480.4636S} {480, 4636}

\bibitem[\protect\citeauthoryear{{Soto}, {Lilly}, {Bacon}, {Richard}  \&
  {Conseil}}{{Soto} et~al.}{2016}]{soto_2016_zap}
{Soto} K.~T.,  {Lilly} S.~J.,  {Bacon} R.,  {Richard} J.,   {Conseil} S.,
  2016, \mn@doi [\mnras] {10.1093/mnras/stw474}, \href
  {https://ui.adsabs.harvard.edu/abs/2016MNRAS.458.3210S} {458, 3210}

\bibitem[\protect\citeauthoryear{Torrey et~al.,}{Torrey
  et~al.}{2015}]{torrey_2015_prog_bias}
Torrey P.,  et~al., 2015, \mn@doi [Monthly Notices of the Royal Astronomical
  Society] {10.1093/mnras/stv1986}, 454, 2770

\bibitem[\protect\citeauthoryear{{Valdes}, {Gupta}, {Rose}, {Singh}  \&
  {Bell}}{{Valdes} et~al.}{2004}]{valdes_2005_indous}
{Valdes} F.,  {Gupta} R.,  {Rose} J.~A.,  {Singh} H.~P.,   {Bell} D.~J.,  2004,
  \mn@doi [\apjs] {10.1086/386343}, \href
  {https://ui.adsabs.harvard.edu/abs/2004ApJS..152..251V} {152, 251}

\bibitem[\protect\citeauthoryear{Vaughan et~al.,}{Vaughan
  et~al.}{2024}]{vaughan_2024}
Vaughan S.~P.,  et~al., 2024, \mn@doi [Monthly Notices of the Royal
  Astronomical Society] {10.1093/mnras/stae409}, 528, 5852

\bibitem[\protect\citeauthoryear{{Veale}, {Ma}, {Greene}, {Thomas},
  {Blakeslee}, {McConnell}, {Walsh}  \& {Ito}}{{Veale}
  et~al.}{2017}]{veale_2017_massive}
{Veale} M.,  {Ma} C.-P.,  {Greene} J.~E.,  {Thomas} J.,  {Blakeslee} J.~P.,
  {McConnell} N.,  {Walsh} J.~L.,   {Ito} J.,  2017, \mn@doi [\mnras]
  {10.1093/mnras/stx1639}, \href
  {https://ui.adsabs.harvard.edu/abs/2017MNRAS.471.1428V} {471, 1428}

\bibitem[\protect\citeauthoryear{Virtanen et~al.,}{Virtanen
  et~al.}{2020}]{2020SciPy-NMeth}
Virtanen P.,  et~al., 2020, \mn@doi [Nature Methods]
  {10.1038/s41592-019-0686-2}, \href {https://rdcu.be/b08Wh} {17, 261}

\bibitem[\protect\citeauthoryear{Wake et~al.,}{Wake et~al.}{2017}]{Wake_2017}
Wake D.~A.,  et~al., 2017, \mn@doi [The Astronomical Journal]
  {10.3847/1538-3881/aa7ecc}, 154, 86

\bibitem[\protect\citeauthoryear{{Wang}, {Cappellari}, {Peng}  \&
  {Graham}}{{Wang} et~al.}{2020}]{wang_2020_kinematic_morphology}
{Wang} B.,  {Cappellari} M.,  {Peng} Y.,   {Graham} M.,  2020, \mn@doi [\mnras]
  {10.1093/mnras/staa1325}, \href
  {https://ui.adsabs.harvard.edu/abs/2020MNRAS.495.1958W} {495, 1958}

\bibitem[\protect\citeauthoryear{{Weilbacher} et~al.,}{{Weilbacher}
  et~al.}{2020}]{weilbacher_2020_data}
{Weilbacher} P.~M.,  et~al., 2020, \mn@doi [\aap]
  {10.1051/0004-6361/202037855}, \href
  {https://ui.adsabs.harvard.edu/abs/2020A&A...641A..28W} {641, A28}

\bibitem[\protect\citeauthoryear{{Wellons} \& {Torrey}}{{Wellons} \&
  {Torrey}}{2017}]{wellons_2017_prog_bias}
{Wellons} S.,  {Torrey} P.,  2017, \mn@doi [\mnras] {10.1093/mnras/stx358},
  \href {https://ui.adsabs.harvard.edu/abs/2017MNRAS.467.3887W} {467, 3887}

\bibitem[\protect\citeauthoryear{Westfall et~al.,}{Westfall
  et~al.}{2019}]{Westfall_2019}
Westfall K.~B.,  et~al., 2019, \mn@doi [The Astronomical Journal]
  {10.3847/1538-3881/ab44a2}, 158, 231

\bibitem[\protect\citeauthoryear{Yan et~al.,}{Yan et~al.}{2015}]{Yan_2016}
Yan R.,  et~al., 2015, \mn@doi [The Astronomical Journal]
  {10.3847/0004-6256/151/1/8}, 151, 8

\bibitem[\protect\citeauthoryear{{van Dokkum} \& {Franx}}{{van Dokkum} \&
  {Franx}}{1996}]{dokkum_1996_fundamental}
{van Dokkum} P.~G.,  {Franx} M.,  1996, \mn@doi [\mnras]
  {10.1093/mnras/281.3.985}, \href
  {https://ui.adsabs.harvard.edu/abs/1996MNRAS.281..985V} {281, 985}

\bibitem[\protect\citeauthoryear{{van de Sande} et~al.,}{{van de Sande}
  et~al.}{2017a}]{jesse_2017_revising}
{van de Sande} J.,  et~al., 2017a, \mn@doi [\mnras] {10.1093/mnras/stx1751},
  \href {https://ui.adsabs.harvard.edu/abs/2017MNRAS.472.1272V} {472, 1272}

\bibitem[\protect\citeauthoryear{{van de Sande} et~al.,}{{van de Sande}
  et~al.}{2017b}]{Jesse_2017_revisiting}
{van de Sande} J.,  et~al., 2017b, \mn@doi [\apj]
  {10.3847/1538-4357/835/1/104}, \href
  {https://ui.adsabs.harvard.edu/abs/2017ApJ...835..104V} {835, 104}

\bibitem[\protect\citeauthoryear{{van der Wel} et~al.,}{{van der Wel}
  et~al.}{2011}]{van_der_wel_2011_majority}
{van der Wel} A.,  et~al., 2011, \mn@doi [\apj] {10.1088/0004-637X/730/1/38},
  \href {https://ui.adsabs.harvard.edu/abs/2011ApJ...730...38V} {730, 38}

\bibitem[\protect\citeauthoryear{van de Sande et~al.,}{van de Sande
  et~al.}{2018}]{jesse_2018_cosmological}
van de Sande J.,  et~al., 2018, \mn@doi [Monthly Notices of the Royal
  Astronomical Society] {10.1093/mnras/sty3506}, 484, 869

\bibitem[\protect\citeauthoryear{van de Sande et~al.,}{van de Sande
  et~al.}{2021}]{jesse_2021_statistical}
van de Sande J.,  et~al., 2021, \mn@doi [Monthly Notices of the Royal
  Astronomical Society] {10.1093/mnras/stab1490}, 505, 3078

\makeatother
\end{thebibliography}

%%%%%%%%%%%%%%%%%%%%%%%%%%%%%%%%%%%%%%%%%%%%%%%%%%

%%%%%%%%%%%%%%%%% APPENDICES %%%%%%%%%%%%%%%%%%%%%
% \appendix
% \section{Some extra material}
%%%%%%%%%%%%%%%%%%%%%%%%%%%%%%%%%%%%%%%%%%%%%%%%%%
% Don't change these lines
\bsp	% typesetting comment
\label{lastpage}
\end{document}